\documentclass{acmsmalltr}
\usepackage{pdfsync}
\usepackage{verbatim}
\usepackage{graphicx}
\usepackage{xspace}
\usepackage{multirow}
\usepackage{rotating}
\usepackage{amsmath}

\def\..{\,\mathpunct{\ldotp\ldotp}} % Middle stuff for intervals. Usage: \..

\newcommand{\Z}{\mathbf Z}

\newcommand{\xorshift}[1][]{\texttt{xorshift#1}\xspace}
\newcommand{\xorshifts}[1][]{\texttt{xorshift#1*}\xspace}
\newcommand{\xorshiftp}[1][]{\texttt{xorshift#1+}\xspace}

\newcommand{\mt}[1][]{\texttt{MT19937}\xspace}
\newcommand{\xorgens}[1][]{\texttt{xorgens#1}\xspace}
\newcommand{\xsadd}{\texttt{XSadd}\xspace}
\newcommand{\xst}[3]{#1, #2, #3}

\newcommand{\la}{\langle}
\newcommand{\ra}{\rangle}

\begin{document}
\markboth{S.~Vigna}{Further scramblings of Marsaglia's \xorshift generators}

\bibliographystyle{ACM-Reference-Format-Journals}

\title{Further scramblings of Marsaglia's \xorshift generators}
\author{Sebastiano Vigna
\affil{Universit\`a degli Studi di Milano, Italy}}

\begin{abstract}
\xorshifts generators are a variant of Marsaglia's \xorshift generators 
that eliminate linear artifacts typical of generators based on $\Z/2\Z$-linear
operations using multiplication by a suitable constant. Shortly
after high-dimensional \xorshifts generators were introduced, Saito and
Matsumoto suggested a different way to eliminate linear artifacts based on addition in
$\Z/2^{32}\Z$, leading to the \xsadd generator. Starting from the
observation that the lower bits of \xsadd are very weak, as
its reverse fails several statistical tests, we explore
variants of \xsadd using 64-bit operations, and describe in
detail \xorshiftp[128], an extremely fast generator that passes
strong statistical tests using only three shifts, four xors and an addition.
\end{abstract}




\acmformat{Sebastiano Vigna, 2014. Further scramblings of Marsaglia's \xorshift
generators.}

\maketitle

\begin{bottomstuff}
Author's addresses: Sebastiano Vigna, Dipartimento di Informatica,
Universit\`a degli Studi di Milano, via Comelico 39, 20135 Milano MI, Italy.
\end{bottomstuff}

\section{Introduction}

\xorshift generators are a simple class of pseudorandom number generators
introduced by~\citeN{MarXR}. While it is known that such generators have some
deficiencies~\cite{PaLXRNG}, the author has shown recently that high-dimensional
\xorshifts generators, which scramble the output of a \xorshift using
multiplication by a constant, pass the strongest statistical tests of the
TestU01 suite~\cite{LESTU01}.

Shortly after the introduction of high-dimensional \xorshifts generators,
\citeN{SaMXA} proposed a different way to eliminate linear artifacts: instead of
multiplying the output of the underlying \xorshift generator (based on $32$-bit shifts) by a
constant, they add it (in $\Z/2^{32}\Z$) with the previous output. Since the sum
in $\Z/2^{32}\Z$ is not linear over $\Z/2\Z$, the result should be
free of linear artifacts.

Their generator \xsadd has $128$ bits of state and full period $2^{128}-1$.
However, while \xsadd passes BigCrush, its \emph{reverse} fails
the LinearComp, MatrixRank, MaxOft and Permutation test of
BigCrush, which highlights a significant weakness in its lower bits.

In this paper, leveraging the theoretical and experimental data about \xorshift
generators contained in~\cite{VigEEMXGS}, we study \xorshiftp, a family of
generators based on the idea of \xsadd, but using 64-bit operations. In
particular, we propose a tightly coded \xorshiftp[128] generator that
does not fail any test from the BigCrush suite of TestU01
(even reversed) and generates 64 pseudorandom bits in $1.06$\,ns on an
Intel\textregistered{} Core\texttrademark{} i7-4770 CPU @3.40GHz (Haswell). It
is the fastest full-period generator we are aware of with such empirical
statistical properties.

% The software used to perform the experiments described in this paper is
% distributed by the author under the GNU General Public License. Moreover,
% all files generated during the experiments are available from the author.

\section{{\fontsize{11pt}{11pt}\selectfont\xorshift} generators}

The basic idea of \xorshift generators is that the state is modified by
applying repeatedly a shift and an exclusive-or (xor) operation. In this paper we consider
64-bit shifts and states made of $2^n$ bits, with $n\geq 7$. We usually
append $n$ to the name of a family of generators when we need to restrict the discussion to a specific state size.

In linear-algebra terms, if $L$ is the $64\times 64$ matrix on $\Z/2\Z$ that
effects a left shift of one position on a binary row vector (i.e., $L$ is all zeroes except for ones on the
principal subdiagonal) and if $R$ is the right-shift
matrix (the transpose of $L$), each left/right shift/xor can be described as a linear multiplication
by $\bigl(I+L^s\bigr)$ or $\bigl(I+R^s\bigr)$, respectively, where $s$ is the
amount of shifting.\footnote{A more detailed study of the linear algebra behind \xorshift generators can be found in~\cite{MarXR,PaLXRNG}.}

As suggested by~\citeN{MarXR}, we use always three low-dimensional
$64$-bit shifts, but locating them in the context of a larger block matrix of the
form\footnote{We remark that \xsadd uses a slightly different matrix, in which the bottom right
element is $1+L^c$.}
\[
M=\left(\begin{matrix}
0 & 0 & 0 & \cdots & 0 & ( I + L^a ) ( I+ R^b )\\
I & 0 & 0 & \cdots & 0 & 0\\
0 & I & 0 & \cdots & 0 & 0\\
0 & 0 & I & \cdots & 0 & 0\\
\cdots&\cdots&\cdots&\cdots&\cdots&\cdots\\
0 & 0 & 0 & \cdots & I & ( I + R^c ) \\
\end{matrix}\right).
\]
It is useful to associate with a linear transformation $M$ its \emph{characteristic polynomial}
\[
P(x)=\operatorname{det}(M-xI).
\]
The associated generator has maximum-length period if and only if $P(x)$ is
primitive over $\Z/2\Z$. This happens if $P(x)$ is irreducible and if $x$ has
maximum period in the ring of polynomial over $\Z/2\Z$ modulo $P(x)$.

The \emph{weight} of $P(x)$ is the number
of terms in $P(x)$, that is, the number of nonzero coefficients. It is considered a good property for generators
of this kind that the weight is close to $n/2$, that is, that the polynomial
is neither too sparse nor too dense~\cite{ComHCRBS}.

\section{{\fontsize{11pt}{11pt}\selectfont\xorshiftp} generators}

It is known that \xorshift generators exhibit a number of linear artifacts,
which results in failures in TestU01 tests like MatrixRank, LinearComp and
HammingIndep. Nonetheless, very little is necessary to eliminate such artifacts:
\citeN{MarXR} suggested multiplication by a constant, which is the approach used
by \xorshifts~\cite{VigEEMXGS}, or combination with an additive \emph{Weyl generator}, which is
the approach used by \citeN{BreSLPRNGUSX} in his \xorgens
generator.

The approach of \xsadd can be thought of as a further simplification of the
Weyl generator idea: instead of keeping track of a separate generator, \xsadd
adds (in $\Z/2^{32}\Z$) consecutive outputs of an underlying \xorshift
generator.
In this way, we introduce a nonlinear operation without
enlarging the state. In practice, this amounts to returning the sum of the currently
updated word and of the lastly updated word of the state. 

\citeN{SaMXA} claim that \xsadd does not fail
any BigCrush test. This is true of the generator, but not of its \emph{reverse}
(i.e., the generator obtained by reversing the bits of the output). Testing the reverse is important because
of the bias towards high bits of TestU01: indeed, the reverse of
\xsadd fails a number of tests, including some that are not due to linear
artifacts, suggesting that it its lower bits are very weak.

We are thus going to study the \xorshiftp family of
generators, which is built on the same idea of \xsadd (returning the sum of consecutive outputs of an
underlying \xorshift generator) but uses $64$-bit shifts and the
high-dimensional transition matrix proposed by Marsaglia. In this way we can leverage the knowledge gathered about
high-dimensional \xorshift generators developed in~\cite{VigEEMXGS}.
 
\subsection{Equidistribution and full period}
\label{sec:ed}

It is known that a \xorshift generator with a state of $n$ bits is
$n/64$-dimensionally equidistributed,\footnote{In this context, a generator with
$n$ bits of state and $t$ output bits is $k$-dimensionally equidistributed if
over the whole output every $k$-tuple of consecutive output values appears $2^{n-t-k}$ times, except
for the zero $k$-tuple, which appears $2^{n-t-k}-1$ times.} and that the
associated \xorshifts generator inherits this property~\cite{VigEEMXGS}. It is easy to show that a slightly weaker property is true of the associated \xorshiftp generator:

\begin{proposition}
\label{prop:eqdist}
If a \xorshift generator is $k$-dimensionally equidistributed, the associated
\xorshiftp generator if $(k-1)$-dimensionally equidistributed.
\end{proposition}
\begin{proof}
Consider a $(k-1)$-tuple $\la t_1,t_2,\ldots,t_{k-1}\ra$. For each possible
value $x_0$, there is exactly one $k$-tuple $\la x_0,x_1,\ldots,x_{k-1}\ra$
such that $x_{i-1}+x_i=t_i$ (the sum is in $\Z/2^{64}\Z$), for $0<i<k$. Thus,
there are exactly $2^{64}$ appearances of the $(k-1)$-tuple $\la
t_1,t_2,\ldots,t_{k-1}\ra$ in the sequence emitted by a \xorshiftp generator
associated with a $k$-dimensionally equidistributed \xorshift generator, with
the exception of the zero $(k-1)$-tuple, for which the appearance associated
with the zero $k$-tuple is missing.
\end{proof}

Note that in general it is impossible to claim $k$-dimensional equidistribution.
Consider the full-period $6$-bit generator that uses $3$-bit shifts with
$a=1$, $b=2$ and $c=1$. As a \xorshift
generator with a $3$-bit output (the lowest bits), it is $2$-dimensionally
equidistributed.
However, it is easy to verify that the sequence of outputs of the associated \xorshiftp generator contains twice the pair
of consecutive $3$-bit values $\la000,000\ra$, so the generator is $1$-, but not
$2$-dimensionally equidistributed.

An immediate consequence is that every individual bit of the generator 
(and thus \textit{a fortiori} the entire output) has full period:

\begin{proposition}
Every bit of a \xorshiftp generator with $n$ bits of state has period
$2^n-1$.
\end{proposition}
\begin{proof}
Since $n\geq 7$, by Proposition~\ref{prop:eqdist} a \xorshiftp generator is at
least $1$-dimensionally equidistributed, and we just have to apply
Proposition~7.1 from~\cite{VigEEMXGS}.
\end{proof}

We remark that, similarly to a \xorshift or \xorshifts\footnote{It should
be remarked that at least the \emph{two} lowest bits of a \xorshifts
generator satisfy a linear recurrence; they become three if the multiplier
is congruent to $1$ modulo $4$, as it happens in~\cite{VigEEMXGS}.}
generator, the lowest bit of a \xorshiftp generator satisfies a linear
recurrence, as on the lowest bit the effect of an addition is the same as
that of a xor.

\subsection{Choosing the shifts}
\label{sec:setup}

\citeN{VigEEMXGS} provides choices of shifts for full-period
generators with $1024$ or $4096$ bits of state. In this paper, however, we want
to explore the idea of \xorshiftp generators with $128$ bits of state to provide
an alternative to \xsadd that is free of its statistical flaws, and faster on
modern $64$-bit CPUs. Finding generators with a small state space, strong
statistical properties and speed comparable with that of a linear congruential
generator is an interesting practical goal.

We thus computed shifts yielding full-period generators;
in particular, we computed all full-period shift triples such that $a$ is coprime
with $b$ and $a+b\leq 64$ (there are $272$ such triples).
We then ran experiments following the protocol used in~\cite{VigEEMXGS},
which we briefly recall. We \emph{sample} generators by executing a battery
of tests from TestU01, a framework for testing pseudorandom number
generators developed by~\citeN{LESTU01}. We start at 100 different seeds that
are equispaced in the state space. For instance, for a 64-bit state we use the seeds $1+i\lfloor
2^{64}/100\rfloor$, $0\leq i<100$. The tests produce a number of statistics, and
we use the number of failed tests as a measure of low quality.

We consider a test failed if its $p$-value is outside of the interval
$[0.001\..0.999]$. This is the interval outside which TestU01 reports a failure by default. 
We call \emph{systematic} a failure
that happens for all seeds. A more detailed discussion of this choice can be found in~\cite{VigEEMXGS}.
Note that we run our tests both on a generator and on its
reverse, that is, on the generator obtained by reversing the
order of the 64 bits returned. The final score is the sum of the number of tests
failed by a generator and its reverse.

We applied a three-stage strategy using SmallCrush, Crush and BigCrush,
which are increasingly stronger test suites from TestU01. We ran SmallCrush on
all $272$ full-period generators just found, isolating $141$ which had less than
$10$ overall failures. We then ran Crush on the latter ones, and finally
BigCrush on the top $10$ results. 

To get an intuition about the relative strength of the two techniques used to
reduce linear artifacts (multiplication by a constant in \xorshifts generators
versus adding outputs in \xorshiftp generators), we also performed the same
tests on \xorshifts[128] generators, and ran BigCrush on the $20$ full-period
triples for \xorshiftp[1024] generators reported in~\cite{VigEEMXGS}.

\section{Results}

In
Table~\ref{tab:BigCrush128plus} we report the results of BigCrush on the ten
best \xorshiftp[128] generators: we show the number of failures of a generator,
of its reverse, their sum, the weight of the associated polynomial and, finally,
systematic failures, if any; it should be compared with
Table~\ref{tab:BigCrush128star8}, which report results for the ten best \xorshifts[128] generators.
In Table~\ref{tab:BigCrush1024plus} we report the same data for the 20
full-period generators identified in~\cite{VigEEMXGS}, which should
be compared with Table~VI therein.

\begin{table}\tbl{\label{tab:BigCrush128plus}Results of BigCrush on the ten
best \xorshiftp[128] generators following Crush.}{%
\renewcommand{\arraystretch}{1.3}
\begin{tabular}{l|rr|r|r|l}
\multirow{2}{*}{$a$, $b$, $c$} & \multicolumn{3}{c|}{Failures}  &
\multirow{2}{*}{Weight} & \multirow{2}{*}{Systematic failures}\\
& \multicolumn{1}{c}{S} & \multicolumn{1}{c|}{R} & \multicolumn{1}{c|}{+} & \\
\hline
\xst{23}{17}{26} & 34 & 30 & 64 & 61 & ---\\
\xst{26}{19}{5} & 31 & 37 & 68 & 53 & ---\\
\xst{23}{18}{5} & 38 & 32 & 70 & 65 & ---\\
\xst{41}{11}{34} & 31 & 39 & 70 & 61& --- \\
\xst{23}{31}{18} & 48 & 34 & 82 & 57 & ---\\
\xst{21}{23}{28} & 53 & 31 & 84 & 47 & ---\\
\xst{21}{16}{37} & 57 & 29 & 86 & 39 & ---\\
\xst{20}{21}{11} & 66 & 32 & 98 & 51 & ---\\
\xst{25}{8}{55} & 48 & 190 & 238 & 51 & BirthdaySpacings\\
\xst{29}{13}{7} & 532 & 593 & 1125 & 57 &
\begin{minipage}{5cm}\raggedright\vspace*{.5em}RandomWalk1C, RandomWalk1H,
RandomWalk1J, RandomWalk1M, RandomWalk1R\vspace*{.5em}\end{minipage}\\
\end{tabular}}
\end{table}

\begin{table}\tbl{\label{tab:BigCrush1024plus}Results of BigCrush on the
\xorshiftp[1024] generators. The last five generators fail systematically a
large number of tests.}{%
\renewcommand{\arraystretch}{1.3}
\begin{tabular}{l|rr|r|r}
\multirow{2}{*}{$a$, $b$, $c$} & \multicolumn{3}{c|}{Failures}  & \multirow{2}{*}{Weight}\\
& \multicolumn{1}{c}{S} & \multicolumn{1}{c|}{R} & \multicolumn{1}{c|}{+} \\
\hline
\xst{16}{23}{30} & 31 & 32 & 63 & 59\\
\xst{31}{11}{30} & 27 & 38 & 65 & 363\\
\xst{10}{11}{61} & 34 & 33 & 67 & 155\\
\xst{40}{11}{31} & 30 & 39 & 69 & 77\\
\xst{9}{14}{41} & 44 & 25 & 69 & 167\\
\xst{10}{9}{63} & 36 & 34 & 70 & 69\\
\xst{31}{33}{37} & 35 & 39 & 74 & 79\\
\xst{41}{7}{29} & 40 & 34 & 74 & 265\\
\xst{15}{16}{19} & 30 & 45 & 75 & 255\\
\xst{27}{13}{46} & 45 & 32 & 77 & 275\\
\xst{9}{5}{60} & 39 & 38 & 77 & 227\\
\xst{22}{7}{48} & 34 & 44 & 78 & 223\\
\xst{7}{16}{55} & 39 & 41 & 80 & 65\\
\xst{25}{8}{15} & 49 & 32 & 81 & 281\\
\xst{31}{10}{27} & 44 & 39 & 83 & 233\\
\xst{3}{26}{35} & 698 & 38 & 736 & 89\\
\xst{2}{11}{61} & 1108 & 34 & 1142 & 81\\
\xst{1}{13}{7} & 1521 & 46 & 1567 & 113\\
\xst{47}{1}{41} & 894 & 819 & 1713 & 99\\
\xst{51}{1}{46} & 890 & 1080 & 1970 & 111\\
\end{tabular}}
\end{table}

\begin{table}\tbl{\label{tab:BigCrush128star8}Results of BigCrush on the ten
best \xorshifts[128] generators following Crush. All generators fail a
MatrixRank test.}{%
\renewcommand{\arraystretch}{1.3}
\begin{tabular}{l|rr|r|r}
\multirow{2}{*}{$a$, $b$, $c$} & \multicolumn{3}{c|}{Failures}  &
\multirow{2}{*}{Weight} \\
& \multicolumn{1}{c}{S} & \multicolumn{1}{c|}{R} & \multicolumn{1}{c|}{+} \\
\hline
\xst{26}{9}{27} & 128 & 124 & 252 & 29\\
\xst{17}{47}{29} & 131 & 126 & 257 & 27\\
\xst{13}{25}{19} & 129 & 130 & 259 & 51\\
\xst{49}{5}{26} & 134 & 128 & 262 & 63\\
\xst{49}{2}{25} & 128 & 135 & 263 & 43\\
\xst{40}{7}{27} & 141 & 129 & 270 & 47\\
\xst{28}{5}{33} & 140 & 131 & 271 & 39\\
\xst{16}{21}{1} & 143 & 132 & 275 & 65\\
\xst{44}{7}{18} & 133 & 153 & 286 & 53\\
\xst{16}{19}{22} & 144 & 143 & 287 & 45\\
\end{tabular}}
\end{table}

% \begin{table}\tbl{\label{tab:BigCrush1024star8}Results of BigCrush on the $20$
% full-period \xorshifts[1024] generators. The last two generators fail
% systematically several tests.}{%
% \renewcommand{\arraystretch}{	1.3}
% \begin{tabular}{l|rr|r|r}
% \multirow{2}{*}{$a$, $b$, $c$} & \multicolumn{3}{c|}{Failures}  & \multirow{2}{*}{Weight}\\
% & \multicolumn{1}{c}{S} & \multicolumn{1}{c|}{R} & \multicolumn{1}{c|}{+} \\
% \hline
% \xst{1}{13}{7} & 28 & 19 & 47 & 113\\
% \xst{3}{26}{35} & 29 & 22 & 51 & 89\\
% \xst{40}{11}{31} & 24 & 33 & 57 & 77\\
% \xst{15}{16}{19} & 30 & 32 & 62 & 255\\
% \xst{22}{7}{48} & 29 & 33 & 62 & 223\\
% \xst{9}{14}{41} & 32 & 30 & 62 & 167\\
% \xst{41}{7}{29} & 25 & 38 & 63 & 265\\
% \xst{31}{11}{30} & 33 & 32 & 65 & 363\\
% \xst{2}{11}{61} & 25 & 41 & 66 & 81\\
% \xst{10}{11}{61} & 42 & 25 & 67 & 155\\
% \xst{7}{16}{55} & 32 & 35 & 67 & 65\\
% \xst{16}{23}{30} & 35 & 34 & 69 & 59\\
% \xst{25}{8}{15} & 25 & 45 & 70 & 281\\
% \xst{27}{13}{46} & 39 & 32 & 71 & 275\\
% \xst{31}{10}{27} & 40 & 32 & 72 & 233\\
% \xst{9}{5}{60} & 40 & 36 & 76 & 227\\
% \xst{31}{33}{37} & 39 & 39 & 78 & 79\\
% \xst{10}{9}{63} & 31 & 49 & 80 & 69\\
% \xst{51}{1}{46} & 60 & 896 & 956 & 111\\
% \xst{47}{1}{41} & 67 & 907 & 974 & 99\\
% \end{tabular}}
% \end{table}

All \xorshifts[128] generators fail the MatrixRank test: with this
state size, multiplication is not able to hide such linear artifacts from
BigCrush. On the other hand, among the best \xorshiftp[128] generators selected
by Crush some non-linear systematic failure appears.

Table~\ref{tab:full01} compares the BigCrush scores of the generators we discussed.
For \xorshiftp[128] we used the triple $23,18,5$ (Figure~\ref{fig:code128}). For
\xorshifts[128] we used the triple $49,5,26$ and for
\xorshiftp[1024]/\xorshifts[1024] the triple $31,11,30$ (the \xorshifts[1024]
generator is the one proposed in~\cite{VigEEMXGS}).

Our choice of triples is based not only on the BigCrush scores and on polynomial
weight, but also on an additional datum: the result of POP
(``$p$-value of $p$-values'') tests. BigCrush generates $254$ $p$-values, each
corresponding to a specific statistics (the same test might generate several
statistics). If the source is perfectly random, and the statistics 
distribution is known exactly, the $p$-values generated at different points of
the state space should appear to be uniformly distributed. 
We can thus test whether this is true for each one of
the $254$ generated values,\footnote{Actually, four $p$-values (two from the LongestHeadRun test and two from the Fourier3 test) have been dropped as
they are based on rather approximate statistics, as documented by the authors of 
TestU01, and thus tend to generate spurious errors.} using a goodness-of-fit
test to get a $p$-value (which is a $p$-value of $p$-values):
NIST~\cite{RSNSTSRPNGCA} suggests the threshold $10^{-4}$ on a $\chi^2$ test on the counts of the $p$-values
falling in the intervals $[k/10\..(k+1)/10)$, $0\leq k<10$; we
used the more stringent value $10^{-3}$ on a Kolmogorov-Smirnov test for the
uniform (continuous) distribution. The triples we suggest for \xorshiftp do
not fail any POP test, and the same happens for the \xorshifts[1024] generator
suggested in~\cite{VigEEMXGS}.

\begin{figure}[ht]
\centering
\begin{verbatim}
#include <stdint.h>

uint64_t s[2];

uint64_t next(void) { 
  uint64_t s1 = s[0];
  const uint64_t s0 = s[1];
  const uint64_t result = s0 + s1;
  s[0] = s0;
  s1 ^= s1 << 23; // a
  s[1] = s1 ^ s0 ^ (s1 >> 18) ^ (s0 >> 5); // b, c
  return result; 
}
\end{verbatim}
\caption{\label{fig:code128}The \xorshiftp[128] generator used in the tests.}
\end{figure}

\begin{figure}[ht]
\centering
\begin{verbatim}
#include <stdint.h>

uint64_t s[16];
int p;

uint64_t next(void) {
  const uint64_t s0 = s[p];
  uint64_t s1 = s[p = (p + 1) & 15];
  const uint64_t result = s0 + s1;
  s1 ^= s1 << 31; // a
  s[p] = s1 ^ s0 ^ (s1 >> 11) ^ (s0 >> 30); // b, c
  return result;
}
\end{verbatim}
\caption{\label{fig:code1024}The \xorshiftp[1024] generator used
in the tests.}
\end{figure}

\section{Jumping ahead}

The simple form of a \xorshift generator makes it trivial to jump ahead quickly
by any number of next-state steps. If $\bm v$ is the current state, we want to
compute $\bm v M^j$ for some $j$. But $M^j$ is always expressible as a
polynomial in $M$ of degree lesser than that of the characteristic polynomial.
To find such a polynomial it suffices to compute $x^j \bmod P(x)$, where $P(x)$
is the characteristic polynomial of $M$. Such a computation can be easily
carried out using standard techniques (quadratures to find $x^{2^k}\bmod P(x)$,
etc.), leaving us with a polynomial $Q(x)$ such that $Q(M)=M^j$. Now, if \[
Q(x)=\sum_{i=0}^n \alpha_ix^i, \] we have \[ \bm v M^j = \bm v Q(M)
=\sum_{i=0}^n \alpha_i\bm vM^i, \] and now $\bm v M^i$ is just the $i$-th state
after the current one. If we known in advance the $\alpha_i$'s, computing $\bm
vM^j$ requires just computing the next state for $n$ times, accumulating by xor
the $i$-th state iff $\alpha_i\neq0$.\footnote{Brent's
\texttt{ranut} generator~\cite{BreURNGS} contains one of the first
applications of this technique.}

In general, one needs to compute the
$\alpha_i$'s for each desired $j$, but the practical usage of this technique is
that of providing subsequences that are guaranteed to be non-overlapping. We can
fix a reasonable jump, for example $2^{64}$ for a \xorshiftp[128] generator, and
store the $\alpha_i$'s for such a jump as a bit mask. Operating the jump is now
entirely trivial, as it requires at most $128$ state changes. In
Figure~\ref{fig:jump} we show the jump function for the
generator of Figure~\ref{fig:code128}. By iterating the jump function, one can
access $2^{64}$ non-overlapping sequences of length $2^{64}$ 
(except for the last one, which will be of length $2^{64}-1$).

\begin{figure}[ht]
\centering
\begin{verbatim}
#include <stdint.h>

void jump(void) {
    static const uint64_t JUMP[] = { 0x8a5cd789635d2dff,
                                     0x121fd2155c472f96 };
	
    uint64_t s0 = 0;
    uint64_t s1 = 0;
    for(int i = 0; i < sizeof JUMP / sizeof *JUMP; i++)
        for(int b = 0; b < 64; b++) {
            if (JUMP[i] & 1ULL << b) {
                s0 ^= s[0];
                s1 ^= s[1];
            }
            next();
        }

   s[0] = s0;
   s[1] = s1;
}
\end{verbatim}
\caption{\label{fig:jump}The jump function for the generator of Figure~\ref{fig:code128}
in C99 code. It is equivalent to $2^{64}$ calls to \texttt{next()}.}
\end{figure}

\subsection{Speed}

Table~\ref{tab:full01} reports the speed of the generators discussed in the
paper and of their \xorshifts counterparts on an an Intel\textregistered{}
Core\texttrademark{} i7-4770 CPU @3.40GHz (Haswell).
We measured the time that is required to emit $64$ bits, so in the
\xsadd case we measure the time required to emit two $32$-bit values.
We used suitable options to keep the compiler from unrolling loops or extracting
loop invariants.

% \begin{table}\tbl{\label{tab:speed}Time to emit a 64-bit integer on an
% Intel\textregistered{} Core\texttrademark{} i7-4770 CPU @3.40GHz (Haswell).}{%
% \renewcommand{\arraystretch}{1.3}
% \begin{tabular}{l|rrr}
% Algorithm & Speed (ns/64 bits)\\
% \hline
% \xorshiftp[128] & $1.12$\\
% \xorshiftp[1024] & $1.34$\\
% \xorshifts[64] & $1.56$\\
% \xorshifts[128] & $1.45$\\
% \xorshifts[1024] & $1.34$\\
% \xsadd			& $2.06$\\
% \end{tabular}}
% \end{table}

The \xorshiftp[128] case is particularly
interesting because we can update the generator paying essentially no cost for the fact that the
state is made of more than $64$ bits: as it is shown in
Figure~\ref{fig:code128}, we just need, while performing an update, to swap the role of the two $64$-bit
words of state when we move them into temporary variables. The resulting code is 
incredibly tight, and, as it can be seen in Table~\ref{tab:full01}, gives rise to the
fastest generator (also because we no longer need to manipulate
the counter that would be necessary to update a \xorshiftp[1024] generator).

\begin{table}\tbl{\label{tab:full01}A comparison of generators.}{%
\renewcommand{\arraystretch}{1.3}
\begin{tabular}{l|r|rr|r|r|l}
\multirow{2}{*}{Algorithm} &\multicolumn{1}{c|}{Speed} &
\multicolumn{3}{c|}{Failures} & \multirow{2}{*}{$W/n$} &
\multirow{2}{*}{Systematic failures}\\
&\multicolumn{1}{c|}{(ns/64\,b)}& \multicolumn{1}{c}{S} & \multicolumn{1}{c|}{R} & \multicolumn{1}{c|}{+}&\\
\hline
\xorshiftp[128] & $1.06$ & 38 & 32 & 70 & $0.50$ & ---\\ % 65/129 = 0.50
\xorshifts[128] &$1.18$& 134 & 128 & 262 & $0.49$ &MatrixRank\\
\xorshiftp[1024] &$1.32$& 27 & 38 & 65 & $0.35$&---\\
\xorshifts[1024]  &$1.34$& 33 & 32 & 65 & $0.35$&---\\
\xsadd			&$2.06$	& 38 & 850 & 888 & $0.10$ &
\begin{minipage}{3cm}\vspace*{.4em}\raggedright LinearComp, MatrixRank, MaxOft,
Permutation\vspace*{.4em}\end{minipage}\\
\end{tabular}}
\end{table}

% \xorshiftp[128] & $1.12$\\
% \xorshiftp[1024] & $1.34$\\
% \xorshifts[64] & $1.56$\\
% \xorshifts[128] & $1.45$\\
% \xorshifts[1024] & $1.34$\\
% \xsadd			& $2.06$\\

\subsection{Escaping zeroland}

\begin{figure}
\centering
\includegraphics[scale=.8]{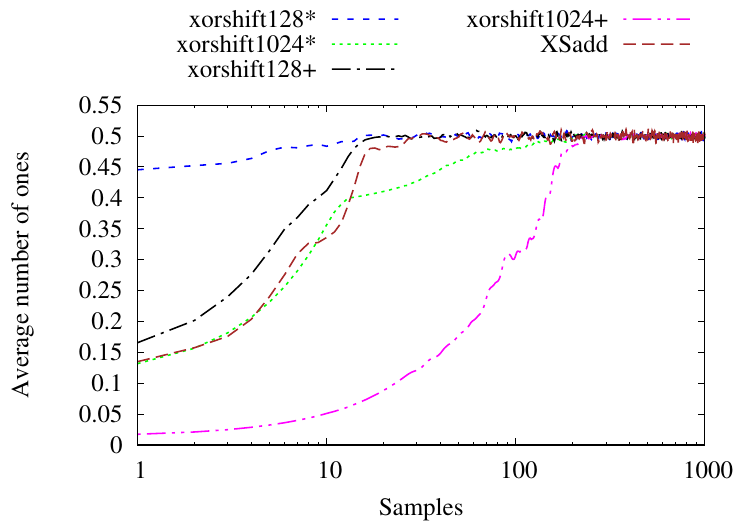}
\caption{\label{fig:ez}Convergence to ``half of the bits are ones in average'' plot.}
\end{figure}

We show in Figure~\ref{fig:ez} the speed at which the generators hitherto
examined ``escape from zeroland''~\cite{PLMILPGBLRM2}:
purely linearly recurrent generators with a very large state space need a very
long time to get from an initial state with a small number of ones to a state in
which the ones are approximately half. The figure shows a measure of escape time
given by the ratio of ones in a window of 4 consecutive 64-bit values sliding
over the first 1000 generated values, averaged over all possible seeds with
exactly one bit set (see~\cite{PLMILPGBLRM2} for a detailed description).
Table~\ref{tab:ez} condenses
Figure~\ref{fig:ez} into the mean and standard deviation of the displayed values.

There are three clearly defined blocks: \xorshifts[128];
then, \xsadd, \xorshiftp[128] and \xorshifts[1024]; finally,
\xorshiftp[1024]. These blocks are reflected also in Table~\ref{tab:ez}. The clear conclusion is
that the \xorshifts approach yields generators with faster escape.

\begin{table}\tbl{\label{tab:ez}Mean and standard deviation for the data shown in Figure~\protect\ref{fig:ez}.}{%
\renewcommand{\arraystretch}{1.3}
\begin{tabular}{l|rr}
Algorithm & Mean & Standard deviation\\
\hline
\xorshifts[128]  & $0.4996$ & $0.0048$ \\
\xorshiftp[128]  & $0.4974$ & $0.0239$ \\
\xsadd			 & $0.4957$ & $0.0302$ \\
\xorshifts[1024] & $0.4935$ & $0.0296$ \\
\xorshiftp[1024] & $0.4575$ & $0.1045$ \\
\end{tabular}}
\end{table}

\section{Conclusions}

We discussed the family of \xorshiftp generators---a variant of
\xsadd based on 64-bit shifts. In particular, we described a \xorshiftp[128]
generator that is currently the fastest full-period generator we are aware of
that does not fail systematically any BigCrush test (not even reversed), making
it an excellent drop-in substitute for the low-dimensional generators found in
many programming languages. For example, the current default pseudorandom
number generator of the Erlang language is a custom \xorshiftp[116] generator
designed by the author using 58-bit integers and shifts (Erlang uses the upper 6
bits for object metadata, so using 64-bit integers would make the algorithm
significantly slower); and the JavaScript engines of Chrome, Firefox and Safari
are based on \xorshiftp[128].
\xorshiftp[128] can also be easily implemented in 
hardware, as it requires just three shift, four xors and an addition.

Higher-dimensional \xorshiftp generators ``escape from zeroland'' too slowly, making them less interesting than their \xorshifts counterpart.

\bibliography{biblio}

\end{document}